\documentclass[apl,noshowpacs,print,twocolumn]{revtex4}
\usepackage{mathrsfs}
\usepackage{graphicx}
\usepackage{dcolumn}
\usepackage{bm}
\usepackage{txfonts}
\newcommand{\upcite}[1]{$^{\mbox{\scriptsize \cite{#1}}}$}

\begin{document}

\title{Spin-filtering and charge- and spin-switching effects in a quantum wire with\\
periodically attached stubs}

\author{Xianbo Xiao,$^{1}$ Zhaoxia Chen,$^2$ Zhiming Rao,$^1$ Wenjie Nie$^1$}
\author{Guanghui Zhou$^{3}$}
\email{ghzhou@hunnu.edu.cn}

\affiliation{$^1$School of Computer, Jiangxi University of
Traditional Chinese Medicine, Nanchang 330004, China}

\affiliation{$^2$School of Mechatronics Engineering, East China
Jiaotong University, Nanchang 330013, China}

\affiliation{$^3$Department of Physics and Key Laboratory for
Low-Dimensional Quantum Structures and Manipulation (Ministry of
Education), Hunan Normal University, Changsha 410081, China}

\begin{abstract}
Spin-dependent electron transport in a periodically stubbed quantum
wire in the presence of Rashba spin-orbit interaction (SOI) is
studied via the nonequilibrium Green's function method combined with
the Landauer-B$\mathrm{\ddot{u}}$ttiker formalism. The coexistence
of spin filtering, charge and spin switching are found in the
considered system. The mechanism of these transport properties is
revealed by analyzing the total charge density and spin-polarized
density distributions in the stubbed quantum wire. Furthermore,
periodic spin-density islands with high polarization are also found inside
the stubs, owing to the interaction between the charge density
islands and the Rashba SOI-induced effective magnetic field. The
proposed nanostructure may be utilized to devise an all-electrical
multifunctional spintronic device.

\end{abstract}

\maketitle

During the last decade, much effort has been made in using the
electron spin in semiconductor to store and communicate information,
often referred to as semiconductor spintronics.\upcite{Awschalom,
Zutic, Fabian} The main challenge in this field is how to provide
feasible methods to generate, manipulate, store, and detect
spin-polarized electrons in semiconductor materials. The Rashba
spin-orbit interaction (SOI),\upcite{Rashba} which arises from the
structural asymmetry of the confining potential in semiconductor
quantum well, is regarded as a promising way. The practical
advantage of the Rashba SOI is that its amplitude can be
conveniently tuned by different means, including the ion
distribution in the nearby doping layers,\upcite{Sherman} the
relative asymmetry of the electron density at the two quantum well
interfaces,\upcite{Golub} and importantly, the applied voltage to
surface gates.\upcite{Engels, Grundler, Koga}

With the advances in modern nanotechnology, various kinds of high
quality low-dimensional nanostructures and superlattice structures
can be fabricated.\upcite{Saminadayar} Recently, spin-dependent
electron transport in nanostructures in the presence of Rashba SOI
has drawn much attention for the purpose of application in
spintronic devices. It is shown that the spin-dependent electron
transport property is very sensitive to the geometrical structure
symmetry of the considered system. For the two-terminal Rashba
nanostructures with broken longitudinal symmetry such as zigzag
wire,\upcite{Zhang2} electron stub waveguide,\upcite{Zhai} and
locally constricted electron waveguides,\upcite{Xiao2, Ban} a
spin-polarized current can be generated in the output lead. In
addition, it has also been found that highly polarized spin-density
islands can be formed inside nonuniform waveguides due to the
structure-induced bound states.\upcite{Zhai,Xiao} For the
two-terminal transversally asymmetrical Rashba nanostructures, such
as hornlike waveguide\upcite{Zhai2} and step-like quantum
wire,\upcite{Xiao3} a very large spin conductance can be obtained
when the forward bias is applied to the structures. However, it is
suppressed strongly when the direction of bias is reversed.

\begin{figure}[b]
\includegraphics[bb=20 34 344 128, width=3.4 in]{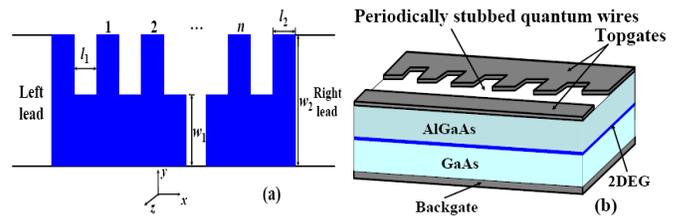}
\renewcommand{\figurename}{FIG.}
\caption{(Color online) Schematic diagrams of (a) a periodically
stubbed quantum wire with Rashba SOI and (b) the experimental setup
proposal, where $l_1$ is the distance between nearest-neighbor stubs
with length $l_2$ and width ($w_2$-$w_1$). The topgates produce the
transverse confinement for the stubbed quantum wire while the
backgate tunes the Rashba SOI strength.}
\end{figure}

Spin-dependent electron transport in superlattice structures has
also been investigated extensively. A periodic and nearly
square-wave pattern with wide gaps has been found in the
transmission spectra of periodically stubbed electron waveguide with
Rashba SOI for a spin-polarized injection, demonstrating the
structure can be utilized to design a spin transistor.\upcite{Wang}
Similarly, transmission gaps with a complete blocking of the charge
current over a range of energy has been found in straight quantum
wires under the modulation of spatially periodic Rashba SOI
(electric field), which can be used as an ideal charge current
switch.\upcite{Wang2,Zhang,Gong,Wang3,Japaridze,Malard,Thorgilsson}
Furthermore, ideal spin-filtering effect can also be achieved in
straight quantum wires when they are modulated by spatially periodic
Rashba SOI and magnetic field simultaneously.\upcite{Wang4,Gong2}
However, spin-dependent electron transport in periodic Rashba
nanostructure with breaking of longitudinal symmetry in the case of
spin-unpolarized injection, up to now, has not been studied yet.

In this letter, by means of nonequilibrium Green's function (GF)
with tight-binding approximation scheme, we calculate the
spin-dependent conductance and local spin-polarized density (LSPD)
distribution for a periodically stubbed quantum wire in the presence
of Rashba SOI. The quantum wire is sandwiched between two normal
metal leads, as sketched in Fig. 1(a), with experimental setup
proposal as shown in Fig. 1(b), where $l_1$ is the distance between
nearest-neighbor stubs with constant length $l_2$ and width
($w_2$-$w_1$). The topgates produce the transverse confinement for
the stubbed quantum wire while the backgate tunes the Rashba SOI
strength. Interestingly, it is found that a spin-polarized current
can be generated in the output lead, which is attributed to the
asymmetrical distribution of spin-polarized density caused by the
Rashba SOI in combination with the broken longitudinal symmetry of
the system. Moreover, ideal charge and spin-polarized current
switching effects are observed in the considered system, owing to
minigaps in the energy band caused by the periodic stub structure.
In addition, highly spin-polarized density islands have also been
found inside stubs due to the interaction between the charge density
islands and the Rashba SOI-induced effective magnetic field.
Therefore, the proposed structure allows for practical applications,
including spin filtering, charge and spin current switching, and
spin storage.

Using the standard tight-binding formalism, the Hamiltonian for the
considered system including Rashba SOI reads
\begin{eqnarray}
H&=&\sum\limits_{lm\zeta}\varepsilon_{lm\zeta}c_{lm\zeta}^{\dag}c_{lm\zeta}-t\sum\limits_{lm\zeta}
\{c_{l+1,m\zeta}^{\dag}c_{lm\zeta}+c_{l,m+1,\zeta}^{\dag}c_{lm\zeta}+\mathrm{H.c.}\}\nonumber\\
&&-\gamma_{R}\sum\limits_{lm\zeta\zeta'}\{c_{l+1,m,
\zeta}^{\dag}c_{lm\zeta'}(i\sigma_{y})^{\zeta\zeta'}-c_{l,m+1,\zeta}^{\dag}
c_{lm\zeta'}(i\sigma_{x})^{\zeta\zeta'}+\mathrm{H.c.}\}\nonumber\\
&&+\sum\limits_{lm\zeta}v_{lm\zeta}c_{lm\zeta}^{\dag}c_{lm\zeta} ,
\end{eqnarray}
where $c_{lm\zeta}^{\dag}(c_{lm\zeta})$ is the creation
(annihilation) operator for an electron with spin $\zeta=\uparrow$
or $\downarrow$ on site $(l, m)$, $\varepsilon_{lm\zeta}=4t$ is the
on-site energy with the electron hopping amplitude
$t=\hbar^{2}/2m^{\ast}a^{2}$ ($m^{\ast}$ and $a$ are the effective
mass of electron and lattice spacing, respectively), $\sigma_{x}$
and $\sigma_{y}$ are the Pauli matrixes,
$\gamma_{R}=\alpha(2a)^{-1}$ is the Rashba SOI strength with the
Rashba coefficient $\alpha$, $v_{lm\zeta}$ is the additional
confining potential, and H.c. means complex conjugate.

According to Hamiltonian (1), the outgoing wave amplitudes and the
wave function of a scattered electron state can be obtained by
adopting the spin-resolved recursive GF method.\upcite{Zhai,Xiao3}
Then the two-terminal spin-dependent conductance is given by the
Landauer-B$\mathrm{\ddot{u}}$ttiker formula\upcite{Buttiker,Pareek}
\begin{eqnarray}
G^{\zeta\zeta'}=\frac{e^2}{h}\mathrm{Tr}(\Gamma_{L}^{\zeta}G_{r}^{\zeta\zeta'}\Gamma_{R}^{\zeta'}G_{a}^{\zeta'\zeta}),
\end{eqnarray}
in which $\Gamma_{L(R)}$ denotes the self-energy function for the
isolated ideal leads, $G_{r}^{\zeta\zeta'}$ $(G_{a}^{\zeta'\zeta})$
is the retarded (advanced) GF of the whole structure, and Tr means
the trace over the spatial and spin degrees of freedom. The total
charge density distribution and the LSPD distribution of the
$z$-component of electrons in the system are defined respectively
as\upcite{Zhai,Brusheim}:
\begin{eqnarray}
\rho^{total}(x,y)=\sum\limits_{q,\zeta}\psi_{q\zeta}^{\dag}(x,y)\psi_{q\zeta}(x,y),
\end{eqnarray}
and
\begin{eqnarray}
LP_{z}(x,y)=\sum\limits_{q,\zeta}\psi_{q\zeta}^{\dag}(x,y)\sigma_{z}\psi_{q\zeta}(x,y),
\end{eqnarray}
where $\psi_{q\zeta}(x,y)$ is the velocity-normalized scattered wave
function of an electron injected from the lead in the $q$th subband
with spin $\zeta=\uparrow$ or $\downarrow$, and the summation is
taken over all the propagating modes in the lead.

The $z$-axis is taken as the spin-quantized axis so that
$|\uparrow>=(1,0)^{\mathrm{T}}$ represents the spin-up state and
$|\downarrow>=(0,1)^{\mathrm{T}}$ denotes the spin-down state, where
T means transposition. Therefore, the charge conductance and the
spin conductance of $z$-component are accordingly defined as
\begin{eqnarray}
G_{e}=G^{\uparrow\uparrow}+G^{\uparrow\downarrow}+G^{\downarrow\downarrow}+G^{\downarrow\uparrow}
\end{eqnarray}
and
\begin{eqnarray}
G_{S_z}=\frac{e}{4\pi}\frac{G^{\uparrow\uparrow}+G^{\downarrow\uparrow}-G^{\downarrow\downarrow}-G^{\uparrow\downarrow}}{e^2/h},
\end{eqnarray}
respectively.

In what follows, we present some numerical examples for the system.
In the calculation the physical quantities are chosen to be that for
a high-mobility GaAs/Al$_{x}$Ga$_{1-x}$
heterostructure\upcite{Yacoby} with a typical electron density
$N\sim 2.5\times 10^{11}$/cm$^2$ and $m^{\ast}=0.067m_{e}$, where
$m_{e}$ is the mass of free electron. The periodically stubbed
quantum wire is formed by adding top-gates upon the heterostructure,
as shown in Fig. 1(b), with fixed geometrical structure parameters
$l_1=l_2=40$ nm, $w_1=36$ nm, and $w_2=80$ nm. The strength of the
Rashba SOI can be controlled by a back-gate and the Rashba
coefficient is set as $\alpha=43.7$ meV$\cdot$nm. For simplicity,
the hard-wall potential approximation is chosen as the transverse
confining potential and the lattice spacing is taken as $a=4$ nm.

\begin{figure}
\center
\includegraphics[bb=29 29 355 179, width=3.4 in]{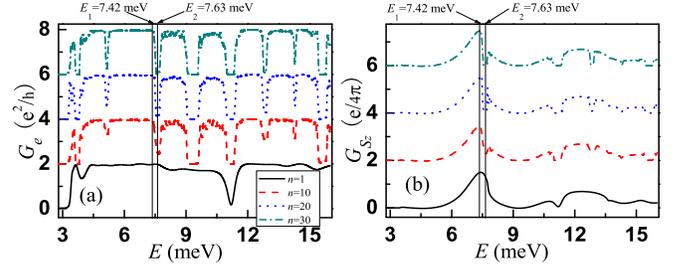}
\renewcommand{\figurename}{FIG.}
\caption{(Color online) The calculated (a) charge conductance and
(b) the $z$-component spin conductance spectra for the system
with number of stubs $n$=1 (black) solid, 10 (red) dashed, 20 (blue)
dotted and 30 (green) dash-dotted, where the graphs are vertically
offset 2.0 for comparison.}
\end{figure}

Figures 2(a) and 2(b) show the charge and spin conductances as
function of the electron energy for various stub
numbers, respectively. For the quantum wire with one stub only, the
resonance and antiresonance structures are found in the charge
conductance, as shown by the (black) solid line in Fig. 2(a),
originating from the interference between the continue states in the
wire and the quasi-bound states in the stub.\upcite{Zhai} When the
number of the stubs increases to $n$=10, minibands appear in the
charge conductance spectrum as shown by the (red) dashed line. As
the number of the stubs increases further, the charge conductance
around the antiresonances decreases gradually and finally extends
into minigaps [see the (blue) dotted and (green) dash-dotted lines
in Fig. 2(a)]. The formation of the minibands and minigaps is
attributed to that the period of the stubs on quantum wire
$L=l_1+l_2=80$ nm is much larger than the lattice spacing $a$=4 nm,
which leads to the split of the energy band. This effect very
resembles the result as found in a Rashba wire under the modulation
of longitudinal periodic potential.\upcite{Thorgilsson} In addition,
the charge conductance oscillations are also found in the allowed
bands between the nearest two minigaps and they become more rapid
with increasing stub numbers. The characteristics of the charge
conductance described above is similar to that of a longitudinal
symmetrical waveguide system with attached stubs.\upcite{Nikolic}
However, the spin conductance exhibits very different behaviors, as
shown in Fig. 2(b). For the quantum wire with single stub, a nonzero
spin conductance can be obtained. Particularly, a broad spin
conductance peak with very large magnitude emerges when the electron
energy at $E_1$=7.42 meV. Interestingly, spin conductance minigaps
appear with the increase of stub numbers and their positions are the
same as that of the charge conductance minigaps. Unexpectedly, the
broad peak happens to be split by a minigap. Moreover, the magnitude
of the spin conductance in the allowed band almost keeps invariable
except for the emergence of some small oscillations, demonstrating
that the spin conductance is mainly determined by the longitudinal
structures while has less relation with the number of stubs.

\begin{figure}
\center
\includegraphics[bb=29 26 355 247, width=3.3 in]{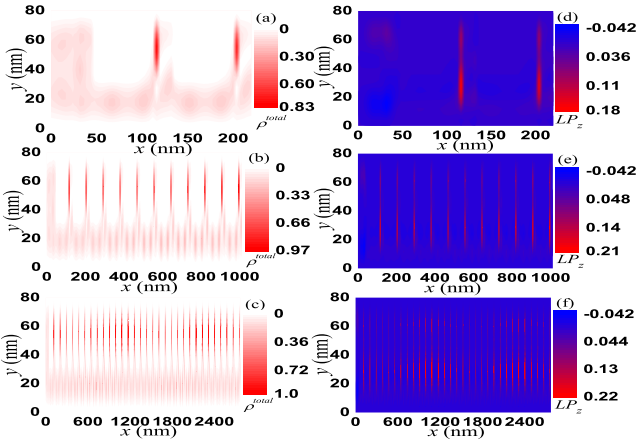}
\renewcommand{\figurename}{FIG.}
\caption{(Color online) The total charge density (the left panels)
and LSPD (the right panels) distributions in stubbed quantum wire
with different length (different number of stubs), where the
electron energy is taken as $E_1=7.42$ meV. The concerned density in
all the panels are normalized against the highest value in (c).}
\end{figure}
\begin{figure}
\center
\includegraphics[bb=29 25 358 249, width=3.3 in]{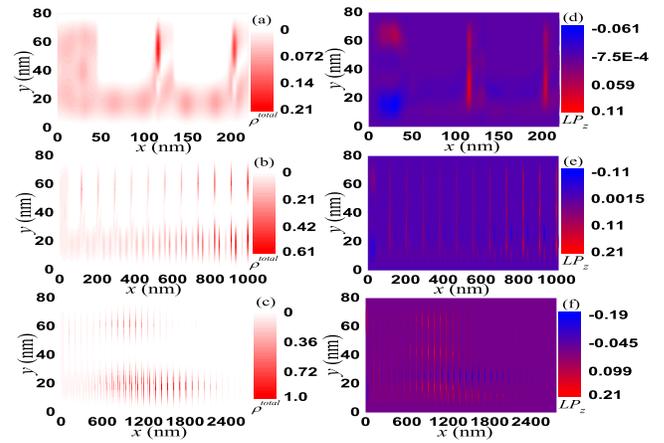}
\renewcommand{\figurename}{FIG.}
\caption{(Color online) The same as Fig. 3 but the energy $E_2$=7.63
meV.}
\end{figure}

In order to further understand the behavior for charge and spin
conductances in Fig. 2, the distributions of total charge density
and LSPD for the system with different number of stubs are shown in
Fig. 3. The electron energy is taken as $E_1$=7.42 meV, i.e., a
value locates inside the broad peak of spin conductance but outside
the minigap of charge conductance (see Fig. 2). For the quantum wire
with single stub, as shown in Fig. 3(a), the total charge
density distribution in the narrow regions and the left wide
region displays regular stripes, which represent propagating modes.
Interestingly, the lower propagating mode spreads along the whole system.
However, the higher propagating mode is strongly localized in the system, that is,
charge density islands are formed inside the stub and the right wide
region. The former island results from the
interference between the forward electron waves and the backward
electron waves scattered by the corners, while the latter island
originates from the interference between the forward electron waves
and the backward electron waves scattered by the wire-lead
interface.\upcite{Xiao3} When the number of the stubs is increased,
as demonstrated in Fig. 3(b) and 3(c), periodic charge density
islands are formed in the stubs and their magnitudes are slightly
enhanced. Meanwhile, the lower propagating mode still spreads along
the whole wire, leading to the unchangeableness of the charge
conductance at this position [see Fig. 2(a)]. Figure 3(d) plots the
LSPD distribution $LP_{z}$ for electrons in the quantum wire with
single stub. Interestingly, highly spin-polarized density islands
are formed inside the stub and the right wide region, owing
to the interaction between the charge density islands and the Rashba
SOI-induced effective magnetic fields.\upcite{Zhai, Xiao} Further,
only spin-polarized density islands with positive sign are generated
inside the stubs because of the longitudinal asymmetry of the
considered system. As a result, a spin conductance with positive
sign can be achieved at this electron energy [see Fig. 2(b)]. In
addition, similar to the behaviors of the total charge density
distribution in Figs. 3(b) and 3(c), periodic spin-polarized density
islands arise in the stubs and their magnitudes strengthened
slightly in some places, as shown in Figs. 3(e) and 3(f), with the
increase of stub numbers. Therefore, the spin conductance within the
allowed band nearly keeps a constant.

Figure 4 shows the total charge density and LSPD distributions of
electrons in the stubbed quantum wire for another energy $E_2=7.63$
meV, a value inside both the broad peak and minigap as seen in Fig.
2. For the quantum wire attached a single stub, as shown in Fig.
4(a), the salient characters of the total charge density
distribution are the same as those in Fig. 3(a). However, the
behaviors of the total charge density distribution for the quantum
wire connected to periodic stubs, as shown in Figs. 4(b) and 4(c),
are very different from that in Figs. 3(b) and 3(c). Apart from the
charge density islands formed inside the stubs, charge density
islands are also formed in the narrow regions of the quantum wire with the increasing
number of stubs. Furthermore, the magnitudes of the charge density
islands are increased dramatically. Finally, the lower propagating mode is also
completely localized in the system, as shown in Fig. 4(c).
Consequently, the charge conductance decreases gradually and reach
zero at final, as seen in Fig. 2(a), with raising stub numbers.
Similarly, the characteristics of the LSPD distribution for the
singly stubbed quantum wires, as depicted in Fig. 4(d), are the same
as that in Fig. 3(d). However, for the quantum wires with periodic
stubs, highly spin-polarized density islands with negative sign
emerge in the narrow areas of the quantum wire besides those inside the stubs with positive
sign, as shown in Figs. 4(e) and 4(f). Moreover, the magnitudes of
the spin-polarized density islands, particularly those with negative
sign, are enhanced obviously and their magnitudes are nearly equal
when the number of stubs is large enough. Thereby, the spin
conductance at this position drops slowly and finally arrives at
zero [see Fig. 2(b)].

In summary, by using the nonequilibrium GF with tight-binding
approximation scheme, we have investigated the spin-dependent
electron transport in a quantum wire with periodically attached
stubs under the modulation of Rashba SOI, where the wire is
connected to two normal metal leads. Our results show that a
spin-polarized current can be achieved in the output lead due to the
longitudinal symmetry of the considered system is broken. Moreover,
both the charge and spin-polarized current can be switched on or off
by varying the electron energy or the number of stubs. In addition,
spin-density islands with high polarization can also be generated inside
the periodic stubs. Therefore, the proposed structure has the
potential for designing multifunctional semiconductor spintronic
devices, such as spin filter, dual charge and spin-polarized current
switch as well as spin memory, without using any magnetic materials
or applying a magnetic field.

This work was supported by the National Natural Science Foundation
of China (Grant Nos. 11264019, 11147156 and 11274108), and by the
Research Foundation of Jiangxi Education Department (Grant No.
GJJ12532).

\end{document}